\documentstyle[12pt,epsf]{article}

\date{\today}

\newcommand\putfig[3]{
   \vbox{
   \let\picnaturalsize=N
   \def\picsize{#3}
   \def\picfilename{#1}
   \ifx\nopictures Y\else{\ifx\epsfloaded Y\else\input epsf \fi
   \let\epsfloaded=Y
   \centerline{\ifx\picnaturalsize N\epsfxsize \picsize\fi
   \epsfbox{\picfilename}}}\fi
   \vspace{1.0cm}
   {\it #2}
   \vspace{1.5cm}
   }
}

\def\be{\begin{equation}}
\def\ee{\end{equation}}
\def\bear{\begin{eqnarray}}
\def\eear{\end{eqnarray}}
\def\nn{\nonumber}

\def\bra{{\langle}}
\def\ket{{\rangle}}
\def\hlf{{{1\over 2}}}

\def\const{{\mbox{const\ }}}                    

\def\lbr{{\lbrack}}
\def\rbr{{\rbrack}}

\def\wdg{{\wedge}}                              

\newcommand\pypx[2]{{{{\partial {#1}}\over{\partial {#2}}}}}

\newcommand\inv[1]{{1\over{#1}}}

\newcommand\rep[1]{{\underline{\bf {#1}}}}      
\newcommand\tr[1]{{\mbox{tr}\{{#1}\}}}          
\newcommand\ev[1]{{\bra {#1} \ket}}             

\newcommand\MS[1]{{{\bf S}^{#1}}}               
\newcommand\MT[1]{{{\bf T}^{#1}}}               

\def\a{{\alpha}}
\def\b{{\beta}}

\def\u{{\mu}}
\def\v{{\nu}}

\def\s{{\sigma}}
\def\t{{\tau}}

\def\tht{{\theta}}

\def\lam{{\lambda}}

\def\Id{{\bf I}}                                

\def\IC{{\bf C}}                                
\def\IZ{{\bf Z}}                                
\def\wdg{{\wedge}}                              

\def\Imx{{\mbox{Im}}}                           
\def\Rex{{\mbox{Re}}}                           

\newcommand\SUSY[1]{{{\cal N}= {#1}}}           

\def\vw{{\vec{w}}}                              

\newcommand\ol[1]{{{\cal O}({#1})}}             
\def\npb#1#2#3{{\it Nucl.\ Phys.} {\bf B#1} (19#2) #3}

\def\prd#1#2#3{{\it Phys.\ Rev.} {\bf D#1} (19#2) #3}

\def\hepth#1{{\it hep-th/{#1}}}

\begin{document}
\begin{titlepage}
\titlepage
\rightline{PUPT-1641}
\rightline{hep-th/9608109}
\rightline{August 15, 1996}
\vskip 1cm
\centerline {{\Large \bf Toroidal Compactification of Heterotic 6D}}
\centerline {{\Large \bf Non-Critical Strings Down to Four Dimensions}}
                 
\vskip 1cm
\centerline {Ori J. Ganor
\footnote{Research supported by a Robert H. Dicke fellowship and
by DOE grant DE-FG02-91ER40671.}}
\vskip 0.5cm
\begin{center}
\em  origa@puhep1.princeton.edu\\
Department of Physics, Jadwin Hall, Princeton University\\
Princeton, NJ 08544, U. S. A.
\end{center}
\vskip 1cm
\abstract{
The low-energy limit of the 6D non-critical string theory with
$N=1$ SUSY and $E_8$ chiral current algebra compactified on 
${\bf T}^2$ is generically an $N=2$ $U(1)$ vector multiplet.
We study the analog of the Seiberg-Witten solution for the
low-energy effective action as a function of $E_8$ Wilson lines
on the compactified torus and the complex modulus of that torus.
The moduli space includes regions where the Seiberg-Witten
curves for $SU(2)$ QCD are recovered as well as regions where
the newly discovered 4D theories with enhanced $E_{6,7,8}$
global symmetries appear.
}
\end{titlepage}


\section{Introduction}
One of the many new challenges that arose in the recent
developments in physics is to obtain a microscopic description
of the 6D non-critical (and sometimes tensionless) strings.

In 6D there are a few kinds of such theories.
The first is the
theory with $\SUSY{2}$ which was discovered in \cite{WitCOM}
and is related to type-IIB on a K3 with an $A_1$ singularity
and which we will refer to as {\em type-II TS-theory}.
The second is the theory with $\SUSY{1}$ and global
$E_8$ symmetry which is related to small 
$E_8$ instantons \cite{GanHan,SWCSD} and which we will  refer to
as {\em Heterotic TS-theory} (or HTS-theory).
There are other, even more mysterious theories which are related
to type-IIB at special values of the coupling constant or F-theory
on a $\IZ_3$ orbifold \cite{SWCSD,WitPMF,MVII}.

In six dimensions, the low-energy of the first two theories is known
(except for the super-conformal tensionless string point) and
is a free tensor multiplet of $\SUSY{2}$ \cite{WitCOM} or
$\SUSY{1}$ according to the theory.

One way to probe a microscopic structure is to compactify the
theory down to lower dimensions.

It was argued in \cite{WitCOM} that compactification of type-II TS-theory
on $\MT{2}$ is related, at scales much below the compactification scale,
to $\SUSY{4}$ SU(2) Yang-Mills theory.
In a recent paper \cite{MeCOMP} we studied some aspects of the
low-energy of type-II theory compactified to 2D on some 4-manifolds.
The main tool there was the relation between type-II on $\MT{2}$ 
and $\SUSY{4}$ Yang-Mills theory which made it possible to read
off the partition function from results of Vafa and Witten \cite{VWSCT}
by a further compactification on $\MT{2}$ (down to zero dimensions).

In order to implement this technique to the heterotic TS-theory
we need to know the low-energy of HTS-theory on $\MT{2}$.
The low-energy will be a function of 9 complex background
parameters which are the $E_8$ boundary conditions (Wilson lines)
along the $\MT{2}$ and the complex structure $\s$ of $\MT{2}$.

At a generic point in the moduli space the 4D low energy is
given by a $U(1)$ vector multiplet which is the dimensional 
reduction of the 6D tensor multiplet 
and so the question arises what is the
Seiberg-Witten elliptic curve that describes that low-energy.

The answer encodes some interesting physical information.
There are regions of moduli space where by appropriate scaling 
of the parameters we can reach $\SUSY{2}$ QCD \cite{SWQCD}
and there are other regions where we reach the newly discovered
4D theories with enhanced $E_{6,7,8}$ global symmetry 
\cite{BDS,SeiIRD,MinNem}. Indeed, the relation between those theories 
and tensionless strings has been anticipated by Seiberg.

The paper is organized as follows:
\begin{enumerate}
\item
Section (2): Generalities regarding the compactification on $\MT{2}$.
\item
Section (3): We derive the Seiberg-Witten curve as a function of
             the modulus of the torus and the $E_8$ Wilson lines.
\item
Section (4): We discuss appropriate scaling limits which reproduce
             the QCD results \cite{SWQCD} for massive $N_f=3$
             (and hence also $N_f = 0,1,2$) as well as the 
             asymptotically non-free $N_f = 8$!
\item
Section (5): Discussion.
\end{enumerate}

\section{Compactification on $\MT{2}$}

To compactify HTS on $\MT{2}$ we need to define a complex
structure $\s$ for the $\MT{2}$ and $E_8$ flat gauge connections
which serve as backgrounds. Let the area of $\MT{2}$ be denoted by
$A$. In the low-energy limit of 4D we will find generically a 
$U(1)$ vector multiplet of $\SUSY{2}$ super-symmetry.
This is because generically only the 6D tensor multiplet is massless
and it gives rise to the $U(1)$ vector-multiplet while all the
other modes are massive.
The low-energy is thus given by \cite{SWYM}:
\be
\inv{4\pi}\Imx\left\lbrack\int d^4\tht a_D(a) \bar{a}
+\int d^2\tht \hlf\pypx{a_D}{a} W_\a W^\a\right\rbrack
\ee
The gauge coupling constant is given by
\be
\t = {i\over {g^2}} + {\theta\over {2\pi}} = \pypx{a_D}{a} 
\ee
Seiberg and Witten expressed $a_D$ and $a$ of $\SUSY{2}$
Yang-Mills theory as a function
of a third complex variable $u$ which was defined in the 
microscopic theory as
\be
u = \hlf\ev{\tr{\phi^2}}
\ee
In our case, we do not know of any microscopic description
of HTS-theory so we do not know whether there exists a 
``natural'' $u$-variable. At present, we can only ask the
question of what is the function $a(\tau)$. Thus, in our case
$u$ will be only an `auxiliary' variable which parameterizes
the moduli space (but given the simple result that we will find,
it will be very tempting to conjecture that $u$ is a {\em true}
microscopic variable).

The functions $a(u)$ and $a_D(u)$  are give by a Seiberg-Witten curve
\cite{SWYM,SWQCD}:
\be
y^2 = x^3 - f(u) x - g(u)
\ee
where the functions $f,g$
depend on $\s$ and on the flat $E_8$ gauge 
connection that we chose. 

\subsection{Geometrical symmetries}

The moduli space of flat $E_8$ gauge connections is given by a pair
of commuting $E_8$ group elements (Wilson lines) in the Cartan sub-algebra
of $E_8$.
We will take a Cartan sub-algebra which is a double cover of
\be
SO(2)^8 \subset SO(16) \stackrel{\IZ_2}{\longleftarrow}
Spin(16) \subset E_8.
\label{subs}
\ee
Let $\MT{2}$ be given by
\be
\IC/\{z \sim z+2,\, z\sim z+2\s\}
\ee
Let $\a_i$ be the charge under the $i$-th $SO(2)$ of
the Wilson loop around the one cycle of the torus (the
one form $0$ to $2$) and let
$\b_i$ be the charge under the $i$-th $SO(2)$ of
the Wilson loop around the other  cycle (from $0$ to $2\s$).
Then we define
\be
w_i = \a_i + \b_i \s
\ee
A point in the moduli space is given by
\be
(w_1,\dots,w_8)
\ee
with the identifications:
\be
(w_1,\dots,w_8)
\sim
(w_1+n_1 + m_1\s,\dots,w_8+n_8+ m_8\s),\qquad
n_i,m_i\in\IZ,\qquad \sum_1^8 n_i \equiv\sum_1^8 m_i \equiv 0 \pmod{2}
\label{dblp}
\ee
The condition of $0\pmod{2}$ is because of the $\IZ_2$ projection
in (\ref{subs}).

On top of that there are further identifications because of
the $E_8$ Weyl group which is generated by
\bear
(w_1,\dots,w_8) &\rightarrow& (w_{\psi{1}}, \dots, w_{\psi{8}}),
\qquad \psi\in S_8                                      \label{weyl1}\\
(w_1,\dots,w_8) &\rightarrow& ((-1)^{\epsilon_1}w_{1},
          \dots, (-1)^{\epsilon_8}w_{8}),
\qquad \sum_1^8 \epsilon_i \equiv 0 \pmod{2}            \label{weyl2}\\
(w_1,\dots,w_8) &\rightarrow& (w_1 - {{\sum_1^8 w_i}\over 4},
 \dots, w_8 - {{\sum_1^8 w_i}\over 4})                  \label{weyl3}
\eear

The general form of the Seiberg-Witten curve that we are looking
for is:
\be
x^3 - f(u;\s,w_1\dots w_8) x - g(u;\s,w_1\dots w_8)
\label{curv1}
\ee
$f$ and $g$ are analytic functions of the variables.
The argument for analyticity in the $w_i$-s is that they 
can be realized as VEVs of scalar components of background fields
(as in similar cases in \cite{SeiPOH}).

The functions $f,g$ are invariant under 
(\ref{dblp},\ref{weyl1}-\ref{weyl3}),
and in particular (\ref{dblp}) implies that $f,g$ are doubly-periodic
in each $w_i$ separately, with periods $2$ and $2\s$.

Under modular transformations
\be
\s\rightarrow -\inv{\s},\qquad w_i\rightarrow{{w_i-2}\over {\s}}
\ee
The functions $f,g$ are, however, not modular invariant but
as we will see, have modular weights $4$ and $6$.

\subsection{The central charges}

As was explained in \cite{SWQCD}, when there are global $U(1)$-s
present in the low-energy, the $\SUSY{2}$ supersymmetry algebra
contains central charges corresponding to those $U(1)$-s.
In \cite{SWQCD}, the global $U(1)$-s where flavor symmetries
and the central charges where
\be
Z = n_e a + n_m a_D + \sum_i S_i {{m_i}\over {\sqrt{2}}}
\ee
where $S_i$ are the $U(1)$ integer charges.
In our case, the global $U(1)$-s are the eight $SO(2)$-s in
the subgroup\footnote{
In what follows, we will be sloppy and write
$SO(16)\subset E_8$ instead of $Spin(16)$.}
 $SO(2)^8\subset E_8$ that is unbroken by the Wilson
lines. The masses $m_i$ should be replaced by the $w_i$-s.
This can be inferred by representing the $w_i$-s again as 
VEVs of background vector-multiplets.
We will also see in section (4) that at the scaling limits that
reproduce QCD the $w_i$-s become the masses.

Thus:
\be
Z = n_e a + n_m a_D + \sum_i S_i {{w_i}\over 2}
\ee
This cannot be the full central charge because this formula is
not periodic in $w_i$. To correct it, we need to add two
more charges corresponding to the momenta around the two
compact $\MT{2}$ directions:
\be
Z = n_e a + n_m a_D + \sum_i S_i {{w_i}\over 2}
  + P_5 + P_6 \s
\label{Z}
\ee
(we can set $R_5 = 2\pi$ where $R_5$ is the size of the $\MT{2}$
cycle from $0$ to $1$).
Now it is clear that the transformation
$w_i\rightarrow w_i + 2 N + 2 M\s$ must be accompanied by
$P_5\rightarrow P_5 - N S_i$ and $P_6\rightarrow P_6 - M S_i$ as
is also clear physically.

Under modular transformations of $\MT{2}$:
\be
\s \rightarrow - \inv{\s}
\ee
we must have
\be
a \rightarrow -\inv{\s} a_D,\qquad
a_D \rightarrow -\inv{\s} a
\ee
The multiplication by $\s$ expresses the fact that
after changing the axes of the $\MT{2}$ we also change
the units from $R_5 = 2\pi$ to $R_6 = 2\pi$ which multiplies
masses by $|\s|$. The interchange between $a$ and $a_D$
is the physical expectation that monopoles and electrons 
are strings wrapped on different cycles of the $\MT{2}$.

\section{The Seiberg-Witten curve}

In this section we will derive the functions $f$ and $g$ in:
\be
y^2 = x^3 - f(u;\s,w_1\dots w_8) x - g(u;\s,w_1\dots w_8)
\label{curv2}
\ee
The plan is to first obtain the degree of $f$ and $g$ as polynomials
in $u$ from what we physically expect from the region
of moduli space where the tension of the strings is large.
This region will also determine the leading coefficients in $f$
and $g$ as a function of $\s$.
Then we will find the relations between the coefficients of
the $u^k$-s in $f$ and $g$ and the external variables
$\s$ and $w_1\dots w_8$ by using (\ref{Z}) and the
identification \cite{SWQCD} of the central charges and the poles of
the 1-form $\lambda$ which determines $a$ and $a_D$
according to \cite{SWYM}:
\be
{{da} \over {du}} = {{d}\over {du}}\oint d\lambda = \oint {{dx}\over y}.
\ee

\subsection{Large $u$}

In 6D, large tension means that the scalar component
of the tensor multiplet is large compared to the inverse of the
area of the compactified torus.
We will assume that in  4D large string tension corresponds to large
$u$.
Thus we are interested in the region of large $u$ and fixed
$q$ and $\vw$ (but not necessarily small).

To first order, we can forget about the strings altogether and
find the dimensional reduction of $B_{\u\v}^{(-)}$ to 4D.
$B_{\u\v}^{(-)}$ is quantized to have integral periods on $\MT{2}$
and this sets the unit of electric charge in 4D. We find that
to first order the 4D coupling constant is
\be
{i\over {g^2}} + {\tht\over {2\pi}} = \s
\ee
In fact, this result is independent of the Wilson lines $\vw$.

Now let $\Phi$ be the tension of the string (which is the super-partner
of $B_{\u\v}^{(-)}$ in 6D) and let
\be
B = \ev{\int_{\MT{2}} B_{\u\v}^{(-)}}
\ee
For large tension the HTS-instanton expansion 
is an expansion in powers of
\be
e^{-\Phi R_5 R_6 - i B}
\ee
From the curve's point of view, the expansion parameter is $\inv{u}$
(for simplicity we take a right-angled $\MT{2}$ with sides
$R_5$ and $R_6$). So, we identify
\be
u = e^{\Phi R_5 R_6 + i B}.
\ee

The mass of a string that is wrapped around $R_6$ is 
approximately (in units where $R_5 = 2\pi$):
\be
M^2 \approx (\Phi R_6)^2 +  \left({B \over {R_5}}\right)^2
    = |\log u|^2
\ee
The term with $B$ appears because when $B$ is turned on it
shifts the boundary conditions around $R_5$ and is like
``fractional momentum''.
We deduce that for large $u$
\be
a \approx \log u,\qquad {{da}\over {du}}\approx \inv{u}
\label{alogu}
\ee
As in \cite{SWQCD} we write down the general curve
which at large $u$ has modular parameter $\s$:
\be
y^2 = x^3 - \inv{4} g_2(\s) \b^2 x u^{2\a} - \inv{4} \b^3 g_3(\s)u^{3\a}
\ee
here $u^{2\a}$ is the leading (highest power in $u$) coefficient
of $x$ and $u^{3\a}$ is the leading free term.
$\a$ and $\b$ will be determined shortly.
\bear
\inv{4}g_2(\s) &=& 15 \pi^{-4}\sum_{m,n\in \IZ_{\ne 0}}\inv{(m\s+n)^4} 
\nn\\
\inv{4}g_3(\s) &=& 35 \pi^{-6}\sum_{m,n\in \IZ_{\ne 0}}\inv{(m\s+n)^6}
\nn
\eear
This curve has
\be
\oint {{dx}\over {y}} =  \inv{2\sqrt{2\b u^\a}}
\ee
comparing to (\ref{alogu}) we find 
\be
\a = 2,\qquad \b = \inv{8}
\ee
so for large $u$:
\be
y^2 = x^3 - (\inv{256} g_2(\s) u^{4} + \ol{u^3}) x
- \inv{2048} g_3(\s)u^{6} + \ol{u^5}
\ee

We conclude that $f$ is a polynomial of degree $4$ and $g$
is of degree $6$ and we have also determined the leading 
coefficients as a function of $\s$.

\subsection{The $E_8$ Wilson lines}
The curve looks like:
\be
y^2 = x^3 - \sum_{k=0}^4 f_k(\s,w_1\dots w_8)u^k x 
      - \sum_{l=0}^6 g_l(\s,w_1\dots w_8)u^l
\label{curv3}
\ee
We have determined $f_4$ and $g_6$ above as a function $\s$
so there are $4+6 = 10$ remaining coefficients to determine.
Out of those, one coefficient can be set by a shift:
\be
u \rightarrow u + \const
\ee
Since we do not know what $u$ is microscopically anyway
this shift does not matter.
It thus seems that we are left with 9 unknown coefficients
but there are only 8 variables $w_1\dots w_8$ (since we have
already fixed $\s$ to determine $f_4$ and $g_6$) so it
seems that we have a puzzle.
However, precisely when the degrees are $4$ and $6$ we can use
a second degree of freedom:
\be
u \rightarrow \b u,\qquad
x \rightarrow \b^2 x, \qquad
y \rightarrow \b^3 y
\ee
which leaves the holomorphic 2-form
\be
du\wdg {{dx}\over {y}}
\ee
invariant and thus will not change the physical low-energy.

It remains to determine the coefficients as a function of the $w_i$-s.
In fact, we will read off the inverse function from the results
of \cite{SWQCD}.
They showed that the central charges $w_i$ in (\ref{Z}) are 
integrals of 
\be
\Omega = du\wdg {{dx}\over {y}}
\ee
over 2-cycles of the total space of the elliptic 
curve fibered over the $u$-plane.
In our case, the total space is precisely the almost Del Pezzo
surface which appears in the F-theory description
of HTS-theory \cite{MVII,MeTST,KMV}.
This is of course no coincidence, as we have learned in the past
year how abstract Seiberg-Witten curves ``come to life''
in string theory \cite{KacVaf}.

This almost Del Pezzo surface has  the $E_8$ lattice as a sub-lattice
of its $H^2(\IZ)$ cohomology \cite{MVII}.
If we denote the 2-cycles which correspond to this sub-lattice
as $e_1\dots e_8$ then they can be chosen so as not to
intersect the fiber at $u=\infty$ where $\Omega$ has a pole
(see appendix (A) of \cite{MeTST}).

Thus, from the curve (\ref{curv3}) we can get two points
on the $E_8$ root lattice
\be
\int_{e_i}\Imx\lbr {du\wdg {{dx}\over {y}}} \rbr,\qquad
\int_{e_i}\Rex\lbr {du\wdg {{dx}\over {y}}} \rbr
\ee
This corresponds in a natural way to the two Wilson lines on the $\MT{2}$
thus completing the map between $w_1\dots w_8,\s$ and 
the coefficients $f_k$ and $g_l$.

\section{QCD at scaling limits}
In this section we will tune the $w_i$, $u$ and $\s$ in such
a way as to reproduce $SU(2)$ Yang-Mills with matter
in appropriate scaling limits, and in particular see that
we can reproduce the curves of \cite{SWQCD}.

To begin with, let us take the Wilson line along the horizontal
direction of the compact $T^2$ (i.e. the real axis) to be
(in the adjoint representation $\rep{248}$ of $E_8$):
\be
W= \left(
\begin{array}{cc}
 \Id_{120\times 120}   &    0    \\
          0            &    -\Id_{128\times 128} \\
\end{array}
\right)
\label{WilE8}
\ee
This is the special Wilson loop that was used in 
\cite{GanHan} and has the property that 
the $E_8$ heterotic string on an $\MS{1}$ of radius $R_5$
is T-dual to the $SO(32)$ heterotic string on $\MS{1}$ of 
radius $\inv{R_5}$ \cite{GinTOR}.
The $SO(32)$ small instanton has been described by Witten
in the low-energy as a field theory \cite{WitSML}.
It is  an $SU(2)$ gauge theory with $\hlf$-hypermultiplets
in the $(\rep{2},\rep{32})$ of $Sp(1)\times SO(32)$.
In our case, $SO(32)$ is broken by the T-dual of $W$
and we end up with 16 $\hlf$-hypermultiplets in $\rep{2}$.

It was explained in \cite{GanHan} how those states can be
derived from the $E_8$ point of view as winding
states of the string.
The special value of $W$ was important because
for all other values of $W$, the T-dual radius does not go to $\infty$
as $R_5\rightarrow 0$.

Now let us compactify on a second direction of radius
$R_6 = R_5\Imx \s$ and turn on $SO(16)\subset E_8$ Wilson loops
as well as an expectation value for the scalar $\phi$ of the 6D
tensor multiplet and for $\int_{\MT{2}} B_{\u\v}^{(-)}$ and 
see what happens after T-duality.

The tensor multiplet expectation values become $SU(2) = Sp(1)$
Wilson loops and we also get $SO(16)$ Wilson loops.

The $SO(16)$ and $Sp(1)$ Wilson loops will generate masses
$m_i$ to the $(\rep{2},\rep{16})$ $\hlf$-hypermultiplets.

The $SU(2)$ gauge coupling is 1 at the 6D string scale \cite{WitSML}
and in 4D it will be of order 
\be
\inv{g^2} = R_2\inv{R_1}= \Imx\s
\ee
at scale $\Imx\s$ (the area of $\MT{2}$ on the $SO(32)$ side)
and it will continue according to the RGE to the low-energy.

Note that we need to start with both $R_6$ and $\inv{R_5}$ large
so that the $\MT{2}$ on the $SO(32)$ side will be large and we could
use the low-energy description of the small instanton.

In the discussion that follows we will ``forget'' about the previous
section and check what are the restrictions that arise just
from the requirement of compatibility with the QCD curves.

\subsection{Perturbing to small masses}
Let's move a little bit away from
the point with the special Wilson lines (\ref{WilE8}).
This will correspond to giving the 16 hyper-multiplets
8 bare masses
\be
m_1\dots m_{8}
\ee
according to
\be
\delta w_i = {{R_5}\over {2\pi}} m_i
\ee
In the future we will set $R_5 = 2\pi$.

\subsection{Small $q,u$ and $m_1\dots m_8$}
To begin with, we take small $q = e^{2\pi i\s}$ 
and $u$ and $set \delta w_i = m_i$
with all the 8 $m_i$ being small.
In this region, as long as $u^4 \gg q$,
the low-energy coupling constant is given by:
\be
e^{2\pi i\t} = {{\prod (u-m_i^2)} \over {u^4}} q
\label{smallqum}
\ee
and the mass of the $W$ boson is approximately
\be
m_W = \hlf\sqrt{2u}
\ee
The corresponding curve has to be of the form:
\bear
y^2 &=&  (x - u + q^{1/2}\psi(u,m,q^{1/2}))
          (x^2 - {{64q}\over {u^2}}\prod (u-m_i^2)
           + q x \chi(u,m,q^{1/2})+ q^{3/2}\xi(u,m,q^{1/2})) \nn\\
&&\label{curvm1}
\eear
where $\psi,\chi,\xi$ are as yet unknown functions.

By shifting $x$ and requiring the RHS to be a polynomial in $u$
we can determine some of the singular (in $u$) parts in $\psi,\chi$ and 
$\xi$. 
After some algebra we find that the curve is of the form:
\bear
y^2 &=& x^3 - \inv{3}u^2 x - {2\over {27}}u^3 
- 16(\prod m_i)(x + \inv{3} u) q^{1/2} \nn\\
&+& (A(x+ \inv{3}u) + 64 T_7(u))q + \ol{q^{3/2}} \nn\\
&&\label{curvsq}
\eear
where
\be
T_7(u) \equiv \inv{u}\left(\prod (u-m_i^2) - \prod m_i^2\right)
\ee
$A$ is an unknown coefficient which depends on $m_1\dots m_8$,
and all the curves that are polynomials in $u$ 
and of the form (\ref{curvm1}) can
be reached from   (\ref{curvsq}) by the changes:
\bear
x &\rightarrow& x + \chi_0(u) + \chi_1(u)q^{1/2} + \chi_2(u) q \nn\\
u &\rightarrow& u + \psi_1(u)q^{1/2} + \psi_2(u) q             \nn\\
&&\label{uxshift}
\eear
where the coefficients  $\chi_0,\chi_1,\psi_1,\psi_2$ are polynomials
in $u$ with an implicit dependence on the masses $m_1\dots m_8$.

The change in $u$ is to be expected since we do not know the precise
relation between the field theoretic $u$ and the HTS-theoretic $u$.

The degree of $g$ in (\ref{curvsq}) is $7$ and not $6$, but by
using (\ref{uxshift}) with 
\bear
\psi_1(u) &=& \pm 24 u^3 + \ol{u^2} \nn\\
\psi_2(u) &=& -288 u^5 + \ol{u^4} \nn
\eear
we can decrease the degree of $g$ to $6$
at the (good) price of increasing the degree of $f$ to $4$.

\subsection{The $N_f = 3$ scaling limit}
Taking $m_4\dots m_8$ small but fixed and
\be
u  = \lam^2 v,\qquad
q = \lam^2 {{\Lambda_3^2}\over {m_4^2\dots m_8^2}} \equiv
 \lam^2 \Lambda^2,
\qquad m_i = \lam\u_i,\,\,\, i=1,2,3
\label{scaling}
\ee
We reach the limit of $N_f = 3$ $SU(2)$  gauge theory with three
masses $\u_1,\u_2,\u_3$, scale $\Lambda$ and moduli parameter $v$.

After rescaling
\be
x\rightarrow \lam^2 x, \qquad y\rightarrow \lam^3 y
\label{scalexy}
\ee
We need to recover the curve of \cite{SWQCD}:
\bear
y^2 &=& x^2 (x-u) - \inv{64}{\Lambda_3}^2 (x-u)^2 
      - \inv{64}(\u_1^2 + \u_2^2 + \u_3^2){\Lambda_3}^2 (x-u) \nn\\
     &+& \inv{4} \u_1 \u_2 \u_3 \Lambda_3 x
 - \inv{64} (\u_1^2 \u_2^2 + \u_2^2\u_3^2 + \u_1^2\u_3^2){\Lambda_3}^2.
\label{SWN3}
\eear

To compare (\ref{curvsq}) to (\ref{SWN3}) it will be easier to shift
\bear
u &\rightarrow& u - 128 C_5 q \nn\\
x &\rightarrow& x -\inv{3}(u + 128 C_5 q) \nn
\eear
where we define
\be
C_k \equiv \sum_{i_1<i_2<\cdots i_k} m_{i_1}^2 m_{i_2}^2 \cdots m_{i_k}^2
\ee
for $k=1\dots 8$.

After the shifts the curve is in the form:
\bear
y^2 &=& x^2 (x-u) - 16 q^{1/2} x \prod m_i
  -64 C_5 (x-u)^2 q \nn\\
&+& (64 u^7 - 64 C_1 u^6 + 64 C_2 u^5 - 64 C_3 u^4 + 64 C_4 u^3
  + 64 C_6 u - 64 C_7 + A x) q + q^{3/2}\psi(u,x,q,m_i)\nn
\eear
substituting (\ref{scaling}-\ref{scalexy}) in (\ref{curvsq}) we find:
\bear
y^2 &=& x^2 (x-u) - 16\u_1 \u_2 \u_3 M^5\Lambda x \prod m_i
  -64 M^{10} (x-u)^2\Lambda^2  \nn\\
&+& 64 M^{10} (\u_1^2 + \u_2^2 + \u_3^2)\Lambda^2 u
 - 64 M^{10}(\u_1^2 \u_2^2 + \u_2^2\u_3^2 + \u_1^2\u_3^2)\Lambda^2 \nn\\
&+& \inv{\lam^2} A x \Lambda^2 
+ \inv{\lam^3}\Lambda^3 \psi(\lam^2 u,\lam^2 x,\lam\Lambda,\lam\u_i, m_j)\nn
\eear
Comparing with (\ref{SWN3}) we find 
\be
\Lambda_3 = - 64 \Lambda \prod_{j=5}^8 m_j
\ee
and
\be
A = -64 C_6 + \ol{\lam^3}
\ee
where $\ol{\lam^3}$ can be made of $\prod m_i$ and $C_7$ and so on.

We also learn that $\psi = \ol{\lam^4}$ and the curve is:
\bear
y^2 &=& x^2 (x-u) - 16 q^{1/2} x \prod m_i
  -64 C_5 (x-u)^2 q \nn\\
&+& 64 (u^7 - C_1 u^6 + C_2 u^5 - C_3 u^4 + C_4 u^3
  - C_6 (x - u) - C_7 ) q 
+ \chi(m_i) x q + q^{3/2}\psi(u,x,q,m_i)\nn\\
&&\label{curvsqN3}
\eear
where $\chi$ and $\psi$ are unknown but restricted as above.

When we shift $u$ and $x$, we can bring it to the form
(\ref{curv3}) where:
\bear
f &=&
\inv{3} u^2 + 16(u^4 + \prod m_i)q^{1/2} 
+(64 C_6 - {{256}\over 3} C_5 u) q + \ol{q^{3/2}}        \nn\\
g &=& 
-{{2}\over {27}}u^3
 -{{16}\over {3}}(u^4 + \prod m_i)u q^{1/2}  \nn\\
&+& \{{{128}\over 3}C_6 u 
- 64 C_1 u^6 + 64 C_2 u^5 - 64 C_3 u^4 + 64 C_4 u^3 \nn\\
&-& {{320}\over 9}C_5 u^2 + {{128}\over 3}C_6 u - 64 C_7   
- 128 (\prod m_i) u^3\} q 
+ \ol{q^{3/2}} \nn
\eear
Here there is still the freedom (\ref{uxshift}) as long as $f,g$ stay
with degrees $4,6$.
Thus, $u$ in the equation above differs from 
$u$ in (\ref{curv3}) by a change of variables.

\section{Discussion}
We have seen that the Coulomb branch of HTS-theory in 4D
is described by a Seiberg-Witten curve with a discriminant of
degree $12$. 
This low-energy is the same as the one discovered in \cite{BDS,SeiIRD}
for a 3-brane near an $E_8$ singularity of a K3 in F-theory
and indeed the structure of the elliptic fibration of (\ref{curv3})
is the same as that of F-theory near an $E_8$ singularity
\cite{VafaFT}.
In fact the relation with tensionless strings was pointed out
in \cite{SeiIRD}.
HTS-theory also has a Higgs branch which in 6D describes
a large $E_8$ instanton. In 4D this Higgs branch emanates
from singularities on the $u$ plane provided that the
Wilson lines are chosen so as to leave a commutant of the $E_8$
which contains $SU(2)$ (so that a large instanton can be
embedded in it).
It is interesting to study the theory compactified to 3D
as well and the relation the dual field theory description
of \cite{IntSei}.

What can we infer from the 4D low-energy curve 
about the microscopic structure of the 6D theory?

The variable $u$ might be related to an expectation value of
a microscopic HTS operator. For this to be so, it is not
necessary for HTS-theory to have local operators. If 
HTS-theory has only operators that are labeled by surfaces
then the 4D operator could be local at $x$ but to correspond be
labeled by the $\MT{2}$ at a 4D point $x$.
It might be worthwhile to study the instanton terms from
strings wrapped on the $\MT{2}$ (as in \cite{WitNON})
and compare them with the $\inv{u}$ expansion.

Other related questions are the compactifications of other
tensionless string theories.
Like $SU(N)$ Yang-Mills theory the HTS-theories have a natural
$N$ which is one plus the instanton number.
We studied the $SU(2)$ theory but the $SU(N)$, i.e. small
$N+1$ instantons that coincide, might be related to 
generalizations of the hyper-elliptic curves of \cite{KLYT,ArgFar}.

There are other $\SUSY{2}$ tensionless string theories in 4D
 \cite{WitPMF,MVII,HanKle} which can be realized as part of 
the spectrum of type-IIB 
on a Calabi-Yau (or F-theory on a 4-fold CY) and thus
the corresponding Seiberg-Witten curves should also 
be calculable.  It might be interesting to study further compactifications
of those tensionless string theories to 2D.


\end{document}